
\input amstex
\input vanilla.sty
\nopagenumbers
\baselineskip 14pt
\pagewidth{6in}
\pageheight{8.5in}
\font\ninerm=cmr9
\font\tenrm=cmr10
\font\docerm=cmr12
\TagsOnRight
\def\wh{\widehat}

\def\ore{\overrightarrow{=}}
\def\ole{\overleftarrow{=}}
\def\om{\omega}
\def\ov{\overline}

{\docerm
\line{\hfil SB/F-93-213}
\vskip 1cm
\centerline{\bf LIGHT-FRONT DYNAMICS OF}
\centerline{\bf MASSIVE VECTOR CHERN-SIMONS GRAVITY}
\vskip 2cm
\centerline{C. Aragone\footnote"*"{\ninerm{Plenary talk
presented at Silarg  8th, Aguas de Lindoia, July 25-30,1993}}, P. J. Arias}
\centerline{\it Departamento de F\'{\i}sica, Universidad Sim\'on Bol\'{\i}var}
\centerline{\it Caracas 10800-A, Venezuela}
\centerline{\it aragone\@ usb.ve;$\ \ \ \ \ \ \ \ $ parias\@ usb.ve;}
\centerline{\it and}
\centerline{A. Khoudeir}
\centerline{\it Departamentode F\'{\i}sica Universidad de los Andes}
\centerline{\it Apartado 5100, M\'erida, Venezuela}
\centerline{\it adel\@ ciens.ula.ve}
\vskip 2cm
\centerline{\bf ABSTRACT}}
\vskip .3cm

{\tenrm
{\narrower\flushpar
We present a second order gravity action which consists of ordinary Einstein
action augmented by a first-order, vector like, Chern-Simons quasi
topological term. This theory is ghost-free and propagates a pure spin-2 mode.
It is diffeomorphism invariant, although its local Lorentz invariance has
been spontaneuosly broken. We perform the light-front (LF) analysis for both
the linearized system and the exact curved model. In constrast to the 2+1
canonical analysis, in the quasi LF coordinates the differential constraints
can be solved. Its solution is presented here.\par}}

\newpage
\docerm
In three dimensions, it has been pointed out by Deser, Jakiw and
Templeton$^{[1]}$ that addition of the tensorial topological Chern-Simons
term $S_{TCS}$ $\sim <\om\partial\om +\om^3>$ to the Einstein action $S_E$
yields a gauge invariant, ghost free, pure spin-2 massive theory. In this
paper we present a softer possibility, which also gives rise to a massive
spin-2 theory. Instead of the tensorial $CS$ term we introduce the vectorial
$CS$ term constructed out of the dreibein variables $e^a=dx^re_r^a$
$$
S_{VCS}\ore\mu(2\kappa^2)^{-1}<e_p^a\epsilon^{prs}\partial_re_{sa}>.\tag 1
$$

Here $\mu$ is the topological mass of the system, $\kappa$ is the 3-d
gravitational costant, $e_{sa}=e_s{}^b\eta_{ba}$, $\eta_{ba}$ is the flat
Lorentz metric ($-$++) and $\epsilon^{prs}$ is the Levi-Civita density,
$\epsilon^{012}=+1$.

This action is diffeomorphism invariant and it is not local Lorentz invariant.
It is topological in its world indices. It can not be regarded as fully
topological because it needs the flat metric $\eta^{ab}$ to be a good
invariant. Neither it is locally conformally invariant. As it happens with the
other two actions we mentioned before, $S_{VCS}$ alone does not contain local
excitations. The full action we postulate here is
$$
S\ore (2\kappa^2)^{-1}<e_{pa}\epsilon^{pmn}R_{mn}{}^a(\om )>+S_{VCS}\ole
S_E+S_{VCS}\tag 2
$$
where $R_{mn}{}^a\ore \partial_m\om_n{}^a-$
$\partial_n\om_m{}^a-\epsilon^a{}_{bc}\om_m{}^
b\om_n{}^c$ is the planar Riemann tensor, $\om^a\ore dx^r\om_r{}^a$ and
$e^a\ore$ $dx^re_r{}^a$ being respectively the affinity and the dreibein
one-forms.

In spite that Einstein action $S_E$ is both local-Lorentz and diffeomorphism
invariant, and the vector-CS term is only diffeomorphism invariant, complexive
action $S$ is just diffeomorphism invariant too.

The situation is similar with massive tensor CS-gravity, which is the sum of
$S_{TCS}-S_E$. This system is not locally conformal invariant due to the
non conformal invariance of the Einstein action. There is a hierarchy of the
local symmetries, starting with the tensorial $CS$ term
$$
S_{TCS}\ore (2\mu\kappa^2)^{-1}<\om_{pa}\epsilon^{pmn}\partial_m
\om_n{}^a-3^{-1}\epsilon^{pmn}\epsilon_{abc}\om_p^a\om_m^b\om_n^c>\tag 3
$$
which is locally conformal, Lorentz, and diffeomorphism invariant. Then it
comes ordinary Einstein action (2) locally Lorentz and diffeomorphism
invariant and finally one has $S_{VCS}$ which is only diffeomorphism
invariant.

Each of these actions alone has non local excitations. However massive
tensorial CS gravity has a pure spin-2 content. Massive vectorial CS gravity,
Eq. 2, will be shown to have a pure spin-2 content too. One might even go a
step further and lose all gauge invariances.

In that case, one ends up with self-dual gravity$^{[2]}$, a first order action
on flat three dimensional Minkowski space having no gauge invariance, and
a ghost-free, pure spin-2 content.

Independent variations of $\om_p{}^a$, $e_p{}^a$ in $S$ yield the standard
torsionless value of $\om_p{}^a$ in terms of the dreibein variables.
$$
{}_3e\om_p{}^a=e_{pb}e_q{}^a\epsilon^{qrs}\partial_re_s{}^b-
2^{-1}e_p{}^ae_{qb}\epsilon^{qrs}\partial_re_s{}^b\tag 4
$$
and the (second order in $e_p{}^a$) field equations
$$
E^{pa}\ore \epsilon^{pmn}R_{mn}{}^a(\om )+2\mu \epsilon^{pmn}\partial_m
e_n{}^a =0.\tag 5
$$
Their associated Bianchi identities read
$$
\partial_pE^{pa}-\epsilon_{bc}{}^a\om_p{}^bE^{pc}+
2\mu \epsilon^{nrs}\om_n{}^a\om_r{}^be_{sb}=0\tag 6
$$

Insertion of $\om_p{}^a$ as given by Eq. 4 into Eq. 5 leads to the second
order field equations which determine the dynamics of the system. The Riemann
tensor has now a source $\sim \mu\epsilon^{pmn}\partial_m e_n{}^a$ which makes
it locally non trivial.

Consequently now we have the possibility of local excitations, as we will show
below. Physical variations of the dreibein variables under small
diffeomorphisms are $\delta e_r{}^a\sim D_r\xi^a$.

In order to understand the physical content of this theory it is convenient to
analyze the associated linearized system, which can be obtained in
straighfoward manner by introducing $e_{pa}=\eta_{pa}+\kappa h_{pa}$,
$\om_p{}^a=\kappa \om_p{}^a$ in action (2). $S$ then becomes in dimensionless
coordinates $x^r\to x^r \mu^{-1}$
$$
S^{MVG}_{lin} =<\omega_p{}^a\epsilon^{prs}\partial_rh_{sa}>-2^{-1}
<\omega_{pa}\omega^{ap}-\omega^2>+2^{-1}<h_p{}^a\epsilon^{prs}
\partial_rh_{sa}>\tag 7
$$

Our light front variables are $u\equiv 2^{-1/2}(x^0-x^2)$,
$v\equiv 2^{-1/2}(x^0-x^2)$, $\epsilon ^{1vu}=+1$, $\eta^{uu}=\eta^{vv}=0$,
$\eta^{uv}=-1$, $\partial_uf =\dot{f}$, $\partial_vf =f'$.
$$\align
\wh{h}_{1v}'\equiv  h_{1v}\ ,\ \wh{\omega}_{1v}'\equiv \omega_{1v}
\ ,\ & \wh{h}_v\equiv h_{vv}''\ ,\ \wh{\omega}_v\equiv
\omega_{vv}''\ ,\ \omega_{11}=\omega_1\ ,\ \omega_{uu}=\omega_u\\
\wh{h}_{v1}'\equiv  h_{1v}\ ,\ \wh{\omega}_{v1}'\equiv \omega_{1v}
\ ,\ & h_{11}=h_1\ ,\ h_{uu}=h_u\tag 8
\endalign
$$

We have in principle 9+9=18 independent variables.

Writing S in terms of the LF componentes of both $h,\omega$ we obtain, through
straightforward calculations:
$$\align
S_{MVG} = & <-h_{1u}'\wh{h}_{v}'^{\cdot}-\wh{h}_{1v}{h}_{vu}'^{\cdot}
-h_1\wh{h}_{v1}'^{\cdot}>+<-\omega_{1u}'\wh{h}_{v}'^{\cdot}-
\wh{\omega}_{1v}{h}_{vu}'^{\cdot}-\omega_1\wh{h}_{v1}'^{\cdot}>+\\
+ & <-h_{1u}'\wh{\omega}_{v}'^{\cdot}-\wh{h}_{1v}{\omega}_{vu}'^{\cdot}
-h_1\wh{\omega}_{v1}'^{\cdot}>-<-\omega_{1}{\omega}_{vu}+
\wh{\omega}_{v1}{\omega}_{1u}'>+\\
+ & <\omega_u\Cal{C}^u+\omega_{uv}\Cal{C}^v+\omega_{u1}\Cal{C}^1+
h_u\Cal{D}^u+h_{uv}\Cal{D}^v+h_{u1}\Cal{D}^1>\tag 9
\endalign
$$

The terms with the time-derivatives constitute the dynamical germ of the
action, the two following terms $\sim \omega^2$ form the LF-generator and the
last 6 terms show the presence of six Lagrange-multipliers associated with
their respective ordinary differential constraints.

They can solved in terms of the six independent intermediate variables:
$\wh{h}_v$, $\wh{h}_{v1}$, $h_{vu}$, $\wh{\omega}_v$, $\wh{\omega}_{v1}$ and
$\omega_{vu}$.

Their solution has the aspect
$$\align
\Cal{C}^u & \sim \wh{h}_{1v}=\partial_1\wh{h}_v+\wh{\omega}_v\ ,
\Cal{C}^v \sim h_{1v}'=\partial_1h_{vu}-\omega_{vu}+\omega_1\ ,
\Cal{C}^1 \sim h_1 =\partial_1\wh{h}_{v1}+\wh{\omega}_{v1},\tag 10\\
\Cal{D}^u & \sim \wh{\omega}_{1v}=(\partial_1-1)\wh{\omega}_v\ ,
\Cal{D}^v \sim \omega_{1u}'=-\omega_1+(\partial_1+1)\omega_{vu}\ ,
\Cal{D}^1 \sim \omega_1=-\wh{\omega}_{1v}+\partial_1\wh{\omega}_{v1}.\tag 11
\endalign
$$

Insertion of these values into $S_{MVG}$ gives an intermediate action which,
in principle, will have six independent variables.

However some miracle happens. the germ of this intermediate action does not
contain neither $\omega_{vu}$ nor $h_{vu}$. Therefore, variations with respect
to them will give additional constraints.

Variations with respect to $\omega_{vu}$ tell us that
$$
\delta S^{reduced}_{MVG}/\delta\omega_{vu}=0\to
\wh{\omega}_{1v}=\wh{\omega}_{v1}\tag 12
$$

In addition we observe that $h_{vu}$ does not enter into this intermediate
action. The final, reduced, unconstrained action achieves the form
$$
S^{unconstrained}_{MVG}=<\wh{\omega}_v'\wh{\omega}_v^\cdot>-
2^{-1}<\wh{\omega}_v[-\partial^2_1+1]\wh{\omega}_v>\tag 13
$$
where we have shifted $\sqrt{2}\wh{\omega}_v\to \omega^{New}_v$.

This is the typical light-front canonical action for a mass $\mu$ excitation
and shows that linearized MVG does indeed propagates one mass-$\mu$
spin-$2^+(2^-)$ excitation.

It is worth pointing out that in spite action (7) has the unique local gauge
invariance $\delta h_{pa}=\partial_p \xi_a$, which oblige to the existence
of the three differential constraints $\Cal{D}^u,\Cal{D}^v,\Cal{D}^1=0$,
we have found additional differential constraints $\Cal{C}^u,\Cal{C}^v,
\Cal{C}^1=0$ which one would expect if a hidden local Lorentz gauge
invariance would hold.

Action (7) breaks the local Lorentz gauge invariance in a very subtle way.

Now we focus on the curved action $S$ defined in Eq. 2. It is convenient
to work with dimensionless variables $x^r\to \kappa^{-2}x^r$, $\omega\to
\kappa^2\omega$,  $\mu \to \kappa^2\mu$ and with quasi-light front
world-coordinates $(u,v,1)$ in  addition to the local Lorentz ones
$(\ov{u},\ov{v},\ov{1})$. (Bars can be  suppresed without risking ambiguity
since neither $e_{pa}$ nor $\omega_{pa}$ are  symmetric).

The dynamical structure of Eq. 9 persists. $S$ can be written in the form:
$$\align
S = & <e_{1u}\dot{\omega}_v-e_1\dot{\omega}_{v1}+e_{1v}\dot{\omega}_{vu}>+
<(\omega_{1u}+\mu e_{1u})\dot{e}_v - (\omega_1+\mu e_1)\dot{e}_{v1}+\\
+ & (\omega_{1v}+\mu e_{1v})\dot{e}_{vu}>+<\omega_u\Cal{C}^u+\omega_{uv}
\Cal{C}^v+\omega_{u1}\Cal{C}^1+e_u\Cal{D}^u+e_{uv}\Cal{D}^v+e_{u1}\Cal{D}^1>.
\tag 14
\endalign
$$
where $\Cal{C}^a,\Cal{D}^b$ constitute the six differential constraints
asocciated with the six Lagrange multipliers $\omega_{ua},e_{ub}$ this
curved, diffeomorphism invariant actions possess. (Again we follow the
conventions stablished in Eq. 8, double indices are reduced to a single
one, i.e. $\omega_{uu}\to \omega_{u}\cdots$).

The six differential constraints respectively read
$$\align
\Cal{C}^u & \equiv  e_{1v}'+\omega_{v1}e_{1v}-\omega_ve_1+e_{v1}\omega_{1v}-e_v
\omega_1-\partial_1e_v=0,\tag 15 \\
\Cal{C}^v & \equiv  e_{1u}'-\omega_{v1}e_{1u}-\omega_{vu}e_1+e_{vu}\omega_1-
\partial_1e_{vu}-e_{v1}\omega_{1u}=0,\tag 16 \\
\Cal{C}^1 & \equiv  -e_1'+e_{1u}\omega_v-e_{1v}\omega_{vu}+e_v\omega_{1u}-
e_{vu}\omega_{1v}+\partial_1e_{v1}=0,\tag 17
\endalign
$$
and
$$\align
\Cal{D}^u & \equiv  \omega_{1v}'+\omega_{v1}\omega_{1v}+\mu e_{1v}'-
\partial_1\omega_v-\omega_1\omega_v-\mu \partial_1e_v=0,\tag 18 \\
\Cal{D}^v & \equiv  \omega_{1u}'-\omega_{v1}\omega_{1u}+\mu e_{1u}'+
\omega_{vu}\omega_1-\partial_1\omega_{vu}-\mu \partial_1e_{vu}=0,\tag 19 \\
\Cal{D}^1 & \equiv  -\omega_1'-\mu e_1'+\omega_v\omega_{1u}-
\omega_{vu}\omega_{1v}+\partial_1\omega_{v1}+
\mu \partial_1e_{v1}=0,\tag 20
\endalign
$$

In order to proceed further we choose the LF algebraic gauge $e_v=0=e_{v1}=\mu
e_{vu}+\omega_{vu}$. The set of constraints (15)$\cdots$(20) can be regarded as
a set  of six linear ordinary differential equations with respect to the
independent  ``space-like" variable $v$, which determines the values of the six
unknown  functions $e_{1v}$, $e_{1u}$, $e_1,\omega_{1v}$, $\omega_{1u}$,
$\omega_1$ in  terms of the ``coefficients" $\omega_{va}$ and the given gauge
conditions.  Introducing the convenient linear operators $$
D^+ \equiv  \omega_v^{-1}(\partial_v+\omega_{v1})\ ,\
D^- \equiv  \omega_{vu}^{-1}(\partial_v-\omega_{v1})\ ,\
D \equiv  \omega_v^{-1}\partial_v\ ,\
D_1 \equiv  \omega_v^{-1}\partial_1\tag 21
$$
and the quantity $a\equiv \omega_{vu}^{1/2}\ \omega_v^{-1/2}$, the complicated
linear system (15)$\cdots$(20) can be separated and solved.
Its solution has the form:
$$\align
e_1 & =D^+e_{1v},\tag 22 \\
e_{1u} & =2^{-1} a D^{-1} a (D^+ a^{-2} D D^+ + D)e_{1v}-\mu^{-1} a D^{-1}D_1
a \omega_v,\tag 23 \\
\omega_1 & = \mu D^- e_{1u} + \mu D^+ e_{1v} + \omega_v D_1
\ln |\omega_{vu}|, \tag 24 \\
\omega_{1u} & = \mu (D D^- -1)e_{1u} + \mu (3D D^+ + a^2)e_{1v} +
D_1 (\ln |\omega_{vu}| - \omega_{v1}), \tag 25 \\
\omega_{1v} & = \mu (a^{-2}D D^+ +1)e_{1v} - \mu a^{-2} e_{1u}.\tag 26
\endalign
$$
where everything is given in terms of $e_{1v}$.

The key sixth-order linear differential equation which determines $e_{1v}$
as a functional of $\omega_{v1}\ ,\ \omega_{vu}\ ,\ \omega_v$ can be shown to
be: $$\align
& [(D^- D D^- +a^{-2}D)aD^{-1}a(D^+a^{-2}DD^+ +D)+6D^-DD^++4a^{-1}Da]e_{1v}=\\
= &2\mu^{-1}(D^-DD^-+a^{-2}D)aD^{-1}D_1a\omega_v-2\mu^{-1}D^-D_1
(\ln |\omega_{uv}|-\omega_{v1})-2\mu^{-1}\omega_vD_1\ln |\omega_{vu}|.\tag 27
\endalign
$$

The final step in this reduction process is to insert in the full action (14)
the expressions (22)$\cdots$(27) in order to obtain a functional expression
which will depend, in principle, upon the three final variables
$\omega_{v1}\ ,\ \omega_{vu}\ ,\ \omega_v$.

Because of the gauge we have chosen, one has been left with
$$
S_{unconstrained}=<e_{1u}\dot{\omega}_v-e_1\dot{\omega}_{v1}-
\mu^{-1}\omega_{1v}\dot{\omega}_{vu}>.\tag 28
$$
Due to the complexity of dealing with the functional solution for
$e_{1v}$  from Eq. 27 we did not go further. Our aim is to show that, if in the
unscontrained action (28) ones makes independent variations of $\omega_{vu}$
one would get the last constraint which determines $\omega_{v1}$ as a function
of $\omega_v$. Then, its introduction into Eq. 28 will yield the exact
LF-nonlinear action for exact VMG in terms of the unique physical excitation
$\omega_v$.

The results one obtains in a more traditional canonical 2+1 analysis of this
theory has been reported in ref [4]. As it is always the case, the complexity
of the differential constraints in the exact case is higher that in the
LF-analysis and explicit functional solutions seem almost impossible to
obtain.

\newpage

\centerline{\bf REFERENCES}
\vskip .3cm

\item{[1]}S. Deser, R. Jackiw and S. Templeton, Ann. of Phys {\bf 140} (1982)
372, (E) {\bf 185} (1988) 406.
\item{[2]}C. Aragone and A. Khoudeir, Phys. Lett. {\bf B173} (1986) 141;
{\it Quantum Mechanics of Fundamentals Systems 1}, ed. C. Teitelboim, Plenum
Press, New York (1988) pp17.
\item{[3]}C. Aragone, Class Q Gravity {\bf 4} (1987) L1.
\item{[4]}C. Aragone, P. J. Arias and A. Khoudeir, {\it Massive Vector
Cherm-Simons  Gravity}, pre\-print SB/F-92-192.
\bye